\documentclass[a4paper,12pt]{article}
\usepackage[utf8]{inputenc}

\usepackage{amsmath,amssymb,theorem}
\usepackage[english]{babel}

\theoremstyle{plain}

\begin{document}

\title{Mostly Sunny: A Forecast of Tomorrow's Power Index Research}

\author{Sascha Kurz\footnote{
Dept. of Mathematics, University of Bayreuth, Germany. E-mail: {sascha.kurz@uni-bayreuth.de}.}
\;\;
Nicola Maaser\footnote{
Dept. of Economics and ZeS, University of Bremen, Germany. E-mail: {maaser@uni-bremen.de}.}
\;\;
Stefan Napel\footnote{
Dept. of Economics, University of Bayreuth, Germany and
Public Choice Research Centre, Turku, Finland. E-mail: {stefan.napel@uni-bayreuth.de}.}
\;\;
Matthias Weber\footnote{
Center for Research in Experimental Economics and Political Decision
Making, University of Amsterdam, and Tinbergen Institute, The Netherlands. E-mail: {m.g.weber@uva.nl}.}}

\date{\today}
\maketitle

\begin{abstract}

\noindent
Power index research has been a very active field in the last
decades. Will this continue or are all the important questions solved? We
argue that there are still many opportunities to conduct useful research
with and on power indices. Positive and normative questions keep calling
for theoretical and empirical attention. Technical and technological
improvements are likely to boost applicability.

  \medskip
  
\end{abstract}

\section{Introduction}
The 750-page tome {\lq\lq}Power, Voting, and Voting Power: 30 Years After{\rq\rq} which
was edited by Holler and Nurmi (2013) demonstrates that the last three
decades of research on power indices have been very productive. Can this
go on? Or, as Manfred J. Holler put it: {\lq\lq}Is there a future to power index
research?{\rq\rq} -- addressing a scientific community that has seen several
protagonists nominally retire of late.

The fact that two of us have only started to do research on power indices
in the 2010s attests to our firm conviction that there is. There exists a
set of diverse topics on which progress can still be made, and will be
made.

The two recent articles on allocating voting weights in two-tier systems
which have been published the most prominently (Barber\'a and Jackson, 2006;
Koriyama et al., 2013) barely mention classical power measures. This may
be regarded as a dark cloud in the sky of power index research. Top
economics journals are concerned first and foremost with the welfare
properties of voting systems. Power comes as a distant second or even
third (behind epistemic concerns). But welfarist approaches to voting,
which focus on measures of success rather than pivotality, can be viewed
as part of power index research defined in a sufficiently liberal way.
Moreover, we see no evidence that voting power faces greater suspicion
from mainstream economists today than in the past.

We are convinced that today's prospects for power index research are no
worse than 30 years ago. Our academic weather forecast is therefore:
mostly sunny! Power index research will have a productive future. The
specific topics which we expect to be addressed can be grouped loosely
into three areas. In Section 2, we focus on the positive analysis of
voting bodies. We then adopt a more normative, design-oriented perspective
in Section 3. A range of technical issues for which progress is likely are
discussed in Section 4. We conclude in Section 5.

\section{Positive Analysis}
Voting is important for the lives of billions of people. It shapes
democratic participation at all levels of legislature and matters for
decision making in boards or committees in the workplace. It also plays a
role in non-governmental organizations, sports associations, and possibly
even the decision on the next family trip (e.g., Darmann et al., 2012). As
soon as voting and collective decision making come with a minimum of
structure, power indices turn out to be useful.

More countries seem to adopt rather than abandon democratic governance
structures, multinational organizations gain importance and
decision-making bodies which use weighted voting evolve or are even newly
created (see, e.g., Belke and Styczynska, 2006, on the Governing Council
of the European Central Bank). Modern communication technology facilitates
the coordination of geographically dispersed actors in associations and
interest groups. Such organizations rely more and more on formal decision
rules compared to consensus over coffee or beer. Reform suggestions for
the most usual suspects for applications of power indices -- the Council of
the EU, the UN Security Council, the Board of Governors of the IMF -- show
no signs of fading. It is hence easy to affirm: the use of power indices
in applied studies will continue. Some type of power index analysis is
necessary in order to discover unevenness of the democratic playing field,
which may be hidden behind vectors of weights, veto rules, thresholds, and
quorums; it is also needed in order to assess rule changes.

We predict that old distinctions and divisions in the literature will lose
importance, however. For instance, there exists a wide spectrum between
(a) puristic \textit{a priori} analysis, which purposely ignores any preference
patterns of the past in favor of the far-reaching independence and
symmetry assumptions that underlie the Penrose-Banzhaf index (PBI) or
Shapley-Shubik index (SSI; see Felsenthal and Machover, 1998, or Laruelle
and Valenciano, 2008a, for definitions and discussion), and (b)
\textit{a posteriori} analysis which places specific voters, say, individual
members of the US Congress or Supreme Court, on locations in a
multi-dimensional policy space in order to identify the critical Senators
or judges for a given decision. Many normative studies of two-tier voting
systems take correlation between members of the same constituency behind
the constitutional veil of ignorance. Why not do the same in positive
analysis of, say, the IMF or EU?\footnote{Kaniovski (2008) has made promising 
progress in this direction.} The {\lq\lq}veil of ignorance{\rq\rq} is the most
prominent motivation for independence and symmetry presumptions. But some
asymmetries other than voting weights are often part of the game. For
instance, some EU members use proportional and others first-past-the-post
systems in order to determine their Council delegates. Some members of the
IMF have preferential trade agreements or even share the same currency,
others not. This deserves to be accounted for. To some extent, power
indices based on games with a priori unions or a restricted communication
structure have always held a middle ground between pure \textit{a priori} and
\textit{a posteriori} analysis  (see Owen, 1977, and Myerson, 1977, for pioneering
work). But we see scope for more. And we predict that increased public
transparency and improved technology for analyzing voting data will create
a bias towards the \textit{a posteriori} end of the range.\footnote{See,  for instance, 
the use by Badinger et al.\ (2013) of web scraping tools that are provided at 
http://api.epdb.eu/ in order to gather a data set of almost 70,000 individual 
voting decisions of EU member states on more than 3,000 proposals.}

Other dichotomies will also very fruitfully be replaced by a more
pluralistic approach. Helpful as binary distinctions like \textit{a priori} and
\textit{a posteriori}, full approval vs. rejection, P-power vs. I-power,
take-it-or-leave-it committees vs. bargaining committees, etc. may be,
they always narrow one's perspective. The attempt, e.g., to delineate the
power to influence a collective decision ({\lq\lq}I-power{\rq\rq} in the for some time
widely followed terminology of Felsenthal and Machover, 1998) from the
power to appropriate the surplus or {\lq\lq}prize{\rq\rq} generated by it ({\lq\lq}P-power{\rq\rq}) is
certainly praiseworthy. But the seemingly crisp juxtaposition blurs the
fact that both are intertwined, i.e., the distinction is fuzzy at best. It
can therefore be highly misleading to base a categorization of available
power indices on it.\footnote{For instance, the PBI is commonly classified 
as a measure of I-power but also captures P-power in some situations (see 
Felsenthal and Machover, 1998, p.\ 45). The SSI is frequently classified 
as a measure of P-power but also captures I-power in relevant contexts 
(see Napel and Widgr\'en, 2008; Kurz et al.\ 2014a).} It also makes a difference whether a decision making
body can only adopt or reject an exogenous proposal (classified as a
{\lq\lq}take-it-or-leave-it committee{\rq\rq} by Laruelle and Valenciano, 2008a) or if
committee members bargain in search of agreement over a set of feasible
alternatives (a {\lq\lq}bargaining committee{\rq\rq} according to Laruelle and
Valenciano). But it makes a similarly big difference whether the proposals
that are fed into a take-it-or-leave-it committee are composed
strategically by an agenda setter who knows committee members' interests
or whether they are truly exogenous; or whether the set of feasible
alternatives that are negotiated in a bargaining committee is binary
(declare independence or not), one-dimensional (tax rates, emission
thresholds) or higher-dimensional (division of a monetary surplus).

With less {\lq\lq}dichotomism{\rq\rq} and a yet more diverse set of tools, future power
index research will be better prepared to analyze the diverse voting
bodies in the field. Ternary voting games (Felsenthal and Machover, 1997)
allow more accurate positive analysis of, say, power in the UN Security
Council; quaternary dichotomous voting rules (Laruelle and Valenciano,
2012) provide yet more flexibility. Still more general frameworks for
measuring power as pivotality or as outcome sensitivity have been
suggested by Bolger (1993) and Napel and Widgr\'en (2004).

The latter framework is suited also to analyzing collective
decision-making in sequential legislative procedures, which involve
strategic interaction between the relevant players. The so-called
{\lq\lq}ordinary legislative procedure{\rq\rq} of the European Union, formerly referred
to as {\lq\lq}codecision procedure{\rq\rq}, has proposals made or amended by three
different voting bodies in several readings and the possibility of
bargaining in a {\lq\lq}conciliation committee{\rq\rq}. Positive analysis of the balance
of power between European Commission, individual members of the Council,
and the European Parliament therefore requires more than, say, a PBI
calculation.\footnote{See Mayer et al.\ (2013) on analysis of the codecision procedure 
for EU28,  and Felsenthal et al.\ (2003, p.\ 490) on the {\lq\lq}informational poverty{\rq\rq} 
of traditional power indices.}

The fact that conventional indices like the PBI or SSI are so much more
convenient to compute has probably biased applied research in their favor
-- to the detriment of more complicated but perhaps more appropriate
methodology. This adverse fate has presumably also affected the nucleolus
of voting games. Montero (2006) has provided a very convincing motivation
for its use as a power measure when bargaining takes place in the shadow
of a voting rule. To our knowledge, however, its application to the EU
Council by Le Breton et al.\ (2012) has been the first and only.
Fortunately, given that we expect progress on the computational ease of
power index research (see Section 4), we predict a brighter future for
both the nucleolus and strategic analyses of voting procedures.

The blunt question {\lq\lq}Which is the right power index?{\rq\rq} has fortunately been
replaced by more subtle ones, asking which of various properties that go
with distinct indices or methods fit a specific application best.
Different members of the community naturally differ in their answers. The
Holler-Packel index (see Holler and Packel, 1983), for instance, is
vigorously advocated by some while others group it under {\lq\lq}minor indices{\rq\rq}
and hold that {\lq\lq}any reasonable measure of a priori voting power \dots must
respect dominance{\rq\rq} (which the Holler-Packel index does not -- see
Felsenthal and Machover, 2005; 1998, p. 245). Many scholars have expressed
a pronounced preference for the PBI over the SSI at workshops and
conferences; others have done the opposite.

This subjectivity and apparent arbitrariness is a cloud in the sky of
power index research, at least from many outsiders' perspective.
Fortunately, the literature has started to address the details of what
constitutes power in which types of voting situations and what is the
predictive value of power indices on a wider empirical basis. So far,
laboratory experiments have been the method of choice. They provide
maximal control over the aspects of a voting situation that determine a
power index's potential value added. Montero et al.\ (2008), for instance,
have conducted an experiment that empirically demonstrates the paradox of
new members, which was a key prediction of power index analysis. Tentative
support for the SSI and PBI has been found by Geller et al.\ (2004). More
experimental power index research can be expected -- someday perhaps even
in the field.

A related area in which future empirical research could be promising is
concerned with people's preferences for different voting systems. Can
preferences for these be explained by the respective distribution of
voting power, as measured by a particular index? How do people trade off
procedural concerns (e.g., for equal swing probabilities) and personal
success propensities? Weber (2014) provides first evidence that subjects
have a preference for voting systems that allocate Shapley-Shubik power to
group representatives proportionally to group size. These systems are
preferred over ones more in line with Penrose's square root rule to an
extent that is not explicable by classic consequentialism.

\section{Normative Analysis}
The increased pluralism predicted for positive analysis has its natural
analogues -- and has in some cases been preceded by developments -- in
normative analysis. We already pointed to an improved account of given
asymmetries in constitutional analysis. If, for instance, it is a
restriction for the design of a two-tier voting system that the considered
population partition must not be changed into constituencies of equal
size, then it is appropriate to also take the reason for this restriction
behind the veil of ignorance. More generally, power index research will do
well to go beyond maximal symmetry and independence of voters.

Investigations of the {\lq\lq}optimal{\rq\rq} design of two-tier voting systems have
branched into numerous different objective functions since the seminal
investigation by Penrose (1946). Equality of voting power or of expected
utility across individuals, maximal welfare under different utilitarian
assumptions, minimal discrepancy between the outcomes of a two-tier vs. a
direct voting system (with {\lq\lq}discrepancy{\rq\rq} operationalized by the
probability of obtaining different outcomes or some notion of average
outcome distance), and minimal discrepancy between weights and induced
voting powers have all been considered.\footnote{This list should still grow. 
Design of two-tier voting systems with epistemic goals or explicit minority 
protection constraints are promising research areas. It is also an open issue 
to cope with multiple normative criteria simultaneously. For instance, equitable 
representation in UNO or IMF can relate to countries'  population sizes but also 
financial and other contributions to the common objective. No single {\lq\lq}optimal 
rule{\rq\rq} may exist; but which rules are Pareto-maximal with respect to a 
given set of criteria?} The great majority of the studies
have, however, remained faithful to Penrose's original binary setup, i.e.,
have considered a collective decision between two exogenously given
alternatives (say, a random legislative proposal vs. the status quo).
Neither voter abstention is considered nor the possibility of three or
more ordered policy alternatives. Also the cases that binary proposals
arise endogenously from strategic agenda setting or from two-party
competition remain to be explored.

We forecast more departures from the conventional binary focus. There are
still few: Laruelle and Valenciano (2008b) and Le Breton et al.\ (2012)
have analyzed delegated bargaining over a simplex of policy alternatives,
i.e., problems of rent division. Maaser and Napel (2007; 2012; 2014) have
used Monte Carlo simulation in order to study influence-based,
majoritarian, and welfarist objective functions in a median voter
environment with an interval of policy options. Asymptotically optimal
assignments of weights in the latter environment have been analytically
characterized by Kurz et al.\ (2014a) for a democratic fairness objective
similar to Penrose's. Because more than two policy alternatives give rise
to population size effects on the distribution of delegate attitudes, it
is surprising that the pattern obtained from binary setups has re-appeared
also for a continuum of alternatives. Namely, optimal weights relate to
the square root of population sizes in case of independent voters but
plain proportionality is called for in case of at least mildly correlated
constituency members. But the cases in between -- with a finite number but
more than two alternatives -- have not been systematically studied so far.
Preliminary computations indicate that the square root finding for
independent and identically distributed (i.i.d.) voter attitudes may
actually break down. Future research will clarify whether famous square
root results are knife-edge not only with respect to their i.i.d.
assumption but perhaps also with regard to allowing only two policy
options.

A one-dimensional interval of alternatives already allows to analyze
economic questions that would otherwise not be covered (e.g., scope of
regulation, spending on climate change mitigation, monetary policy); it
would be desirable to extend the analysis to multidimensional spaces.
Future research in this vein will have to deal with the {\lq\lq}curse of
multidimensionality{\rq\rq}. One possibility could be to use point solutions,
like the Copeland winner, which exist even if the generalized median voter
does not. Another possibility is to assume an exogenous ordering of
dimensions on which individuals vote sequentially (see De Donder et al.,
2012).

So far, power index research and its normative applications to
representative democracy have stayed closely in the tracks of
winner-takes-all systems, which are easily modeled by weighted voting
games. Other democratic systems like proportional rule or mixed-member
systems have been neglected. We forecast that this will change. Edelman
(2004), for instance, has considered the ideal composition of a
legislature that contains representatives from equipopulous districts and
some number of at-large representatives if the objective is to maximize
the total Banzhaf power of individual citizens. Other scenarios with two
(or even more) types of legislators, representing different interests of
the electorate, are conceivable and will be studied in the future. What,
for instance, should a mixed-member legislature or a two-chamber
legislature ideally look like if voters have interests along ethnic and
economic dimensions, which can be either independent or aligned in
complicated ways?

\section{Tools and Technical Issues}
As in research more generally, the types of power investigations carried
out depend on the available mathematical and computational tools.
Substantial progress has been made regarding the efficient computation of
power indices. Free software packages make it easy to calculate power
indices for applied researchers who do not want to write their own
programs; it is possible to adapt published code to a specific application
(see, e.g., Mac\'e and Treibich , 2012).

Understandably, the availability of software is biased towards the most
popular conventional indices, namely the PBI and the SSI. But popularity
is also a consequence of availability. We are unaware, for example, of any
online tool which would allow an applied researcher to compute the
nucleolus.  For a 27-member assembly, as considered by Le Breton et al.\
(2012), its computation is an almost insurmountable obstacle for
non-experts. So we see a future for more easy-to-use software, especially
for the computation of technically more demanding constructs (as, e.g.,
also the minimum sum representation index recently introduced by Freixas
and Kaniovski, 2014). For power analysis based on convex policy spaces,
algorithmic considerations are still in their infancy.

There is room for improvements even in the computation of SSI and PBI.
Namely, the efficiency of the most widely used generating function
approach (see Alonso-Meijide et al., 2012) relies heavily on working with
small integer weights. This is in stark contrast with population figures
in the millions being used as weights in the EU's Council. Large weights
can also arise when trying to implement Penrose's square root rule as well
as possible. Techniques have recently been developed to compute equivalent
representations with smaller or even the minimum integer weights (see,
e.g., Kurz, 2012a). These may in the future prove worthwhile for index
computations, too.

Another important technical issue is the so-called {\lq\lq}inverse problem{\rq\rq} of
power indices: for a given target distribution of power according to, say,
the PBI or the SSI, one seeks to find a voting rule which induces this
distribution as closely as possible for a given notion of distance. If one
does not want to rely on simple heuristics, which mostly lack provable
qualities such as a known maximal distance to the optimal solution, the
problem is computationally very expensive (see De et al., 2012, and Kurz,
2012b). Progress can still be made regarding a better understanding of
common heuristics (Kurz and Napel, 2014) and regarding the efficient --
ideally also user-friendly -- implementation of exact algorithms. The
usefulness of, e.g., the integer linear programming techniques employed by
Kurz (2012b) will benefit from steadily improving computer hardware; it is
also conceivable that the complete list of distinct weighted voting games
with up to nine players will in coming years become searchable online.

We also forecast progress in the pure theory of power indices. The
distribution of inducible power vectors within the unit simplex is, also
for the classical PBI or SSI, more mystery than understood. In a seminal
recent paper, Alon and Edelman (2010) have shown that even for large
numbers of players some target PBI distributions can be reached only with
a large and constant relative error. Their work is in the process of being
extended to other power indices (see Kurz, 2014).

Another theoretical issue of practical relevance is the possible
coincidence of voting weights and power -- either in an exact or asymptotic
sense. It was shown only recently that the nucleolus of non-oceanic
weighted majority games converges to the relative weight distribution (see
Kurz et al., 2014b). The same article provided a new sufficient condition
for exact coincidence of nucleolus and weights, which future research can
presumably weaken. Coincidence of power and weights has also been studied
recently by Houy and Zwicker (2014) for the PBI. Analogous findings for
the SSI remain to be developed. The first attempt by Leech (2013) to
develop an asymptotic result for power indices which covers both oceanic
and non-oceanic games has turned out to misstate rather than generalize
findings by Lindner and Machover (2004). But the goal was worthwhile, and
we forecast that it will be achieved in future research.

\section{Concluding Remarks}
Above selection of topics for which we expect power index research to
remain fruitful is biased by our own curiosities. That the collection is
obviously too big an agenda for us alone, however, indicates the wide
scope for continuing with or moving into power index research.

This scope becomes even wider if one also considers topics that are more
distantly related to voting power. For instance, the quantifications of
causal responsibility by Braham and van Hees (2009) or Felsenthal and
Machover (2009) draw more or less explicitly on power analysis of
non-strategic binary voting. Carrying methods and insights from non-binary
strategic voting over into this domain looks promising. The domain of
conventional power index research has also been left by Koster et al.'s
(2014) investigation of the predictive value of knowing an individual
voter's decision or voting inclination. Taking the latter as input into a
model of an opinion formation process could merge traditional power
analysis with the analysis of social dynamics and networks.

Finally, indices and techniques that have been popularized by voting
applications can prove useful in completely unrelated contexts. For
example, Kovacic and Zoli (2013) compute the PBI with relative population
shares of different ethnicities as {\lq\lq}weights{\rq\rq} in an analysis of ethnic
conflict.  They find that a PBI-based approach can explain onset of
conflict better than using existing indices of ethnic diversity.

\end{document}